\documentclass[twocolumn,3p]{elsarticle}  

\usepackage{fancyhdr}
 \usepackage{amsmath}
\usepackage{cases}
\usepackage{stfloats}
\usepackage{enumerate}
\usepackage{url}
\usepackage{multicol}
\usepackage{multirow}
\usepackage{graphicx}
\usepackage{wrapfig}
\usepackage{picinpar}
\usepackage{cutwin}
\usepackage{picins}
\usepackage{balance}
\usepackage{algorithm}
\usepackage{algpseudocode}

 \usepackage{etoolbox}

\newtoggle{ACM}
\toggletrue{ACM}
\togglefalse{ACM}

\newtoggle{IEEEcls}
\toggletrue{IEEEcls}
\togglefalse{IEEEcls}

\newtoggle{ElsJ}
\toggletrue{ElsJ}

\floatname{algorithm}{Algorithm}

\usepackage{amsthm}
\theoremstyle{plain}

\theoremstyle{definition}
\newtheorem{definition}{Definition}

\newif\ifNotUse  
 \NotUsetrue
 
 \newif\ifNotUse  
 \NotUsetrue

\begin{document}
\title{Cardinalities estimation under sliding time window by sharing HyperLogLog Counter}
\author[seu_cs]{Jie Xu\corref{cor1}}
\ead{xujieip@163.com}


\begin{abstract}
Cardinalities estimation is an important research topic in network management and security. How to solve this problem under sliding time window is a hot topic. HyperLogLog is a memory efficient algorithm work under a fixed time window. A sliding version of HyperLogLog can work under sliding time window by replacing every counter of HyperLogLog with a list of feature possible maxim (LFPM). But LFPM is a dynamic structure whose size is variable at running time. This paper proposes a novel counter for HyperLogLog which consumes smaller size of memory than that of LFPM. Our counter is called bit distance recorder BDR, because it maintains the distance of every left most ``1" bit position. The size of BDR is fixed. Based on BDR, we design a multi hosts' cardinalities estimation algorithm under sliding time window, virtual bit distance recorder VBDR. VBDR allocate a virtual vector of BDR for every host and every physical BDR is shared by several hosts to improve the memory usage. After a small modifcation, we propose another two parallel versions of VBDR which can run on GPU to handle high speed traffic. One of these parallel VBDR is fast in IP pair scanning and the other one is memory efficient. BDR is also suitable for other cardinality estimation algorithms such as PCSA, LogLog. 
\end{abstract}

 \maketitle
 

\section{Introduction}
Generally speaking, cardinality is the distinct number of elements in a data stream. It is a crucial attribute in network management and research\cite{DosC:ACooperativeIntrusionDetectionFrameworkCloud}\cite{DosC:AnalysisSimulationDDOSAttackCloud}\cite{Scan:EvasionResistantNetworkScanDetection}. In network field, cardinality maybe flows cardinality(the number of distinct flows in a particular period\cite{HSD:sampleFlowDistributionEstimate}), host cardinality(the number of other hosts contacting with it\cite{HSD:streamingAlgorithmFastDetectionSuperspreaders}). Cardinality estimating algorithm could be applied to all these kind of problems. For a brief description, cardinality in this paper means host's cardinality. 

\begin{figure}[!ht]
\centering
\includegraphics[width=0.47\textwidth]{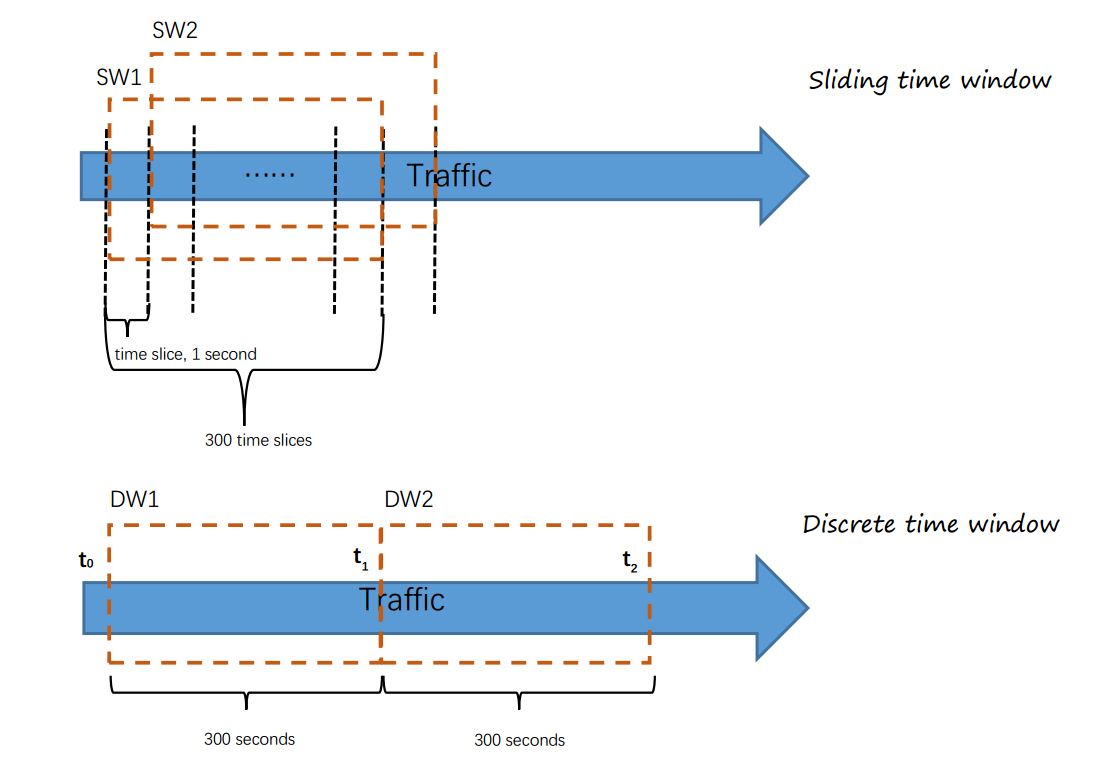}
\caption{Sliding time window and discrete time window}
\label{fig_sliding_and_discrete_time_window}
\end{figure}

The cardinality of a host $aip$ is the number of other hosts communicating with it through an edge router $ER$ in a particular period\cite{HSD:AcontinuousVirtualVectorBasedAlgorithmMeasuringCardinalityDistribution}. We call these hosts, which communicate with $aip$ in a certain period, as aip's opposite hosts.
The period could be a discrete time window or a sliding time window as shown in fig.\ref{fig_sliding_and_discrete_time_window}. The traffic is divided into regular time slices. The size of a time slice could set to 1 second, 1 minute, 5 minutes or any fix duration of different applications.

The time window moves forward a time slice once a time. The sliding time window contains k successive time slices\cite{SDC:IMC2003:IdentifyingFrequentItemsSlidingWindowsOnlinePacketStreams}, but the discrete time window only has a time slice. Let $W(t,k)$ represent a time window starting from the $t$th time slice with $k$ time slices. The size of a time window is the duration it covers. For a sliding time window, its size is the sum duration of the k time slices. For a discrete time window, its size is the duration of a time slice. When the size of a sliding time window equals to that of a discrete time window, the size of a time slice in the sliding time window is $k$ times smaller than that in the discrete time window. The small size of the time slice let the cardinality acquired under sliding time window is more accurate and prompt than that under a discrete time window.

But calculating cardinality under sliding time window is much more complex than that under discrete time window because it needs to preserve the host state of previous time slices when the sliding time window moves forward\cite{SHSD:DistributedStreamsAlgorithmsSlidingWindows}. Let $OP(aip, t, k)$ represent the set of aip's opposite hosts in $W(t, k)$. If $bip \in OP(aip, t, k)$, we say that $bip$ is active for aip in $W(t, k)$. The primary problem of cardinality estimating under sliding time window is to determine if a $bip$ in $OP(aip, t, k)$ is still active in $W(t+1, k)$ at the beginning of time slice $t+k$. For example, if $bip$ appears in $W(t+1, k-1)$, it will be still active at the beginning of time slice $t+k$; if $bip$ only appears in time slice $t$, it will be inactive at the beginning of time slice $t+k$. Sliding time window algorithm must distinguish active and inactive opposite hosts in every time slices. We solve this problem by designing a novel counter, Bit Distance Recorder(BDR). We use BDR as the counter of HyperLogLog\cite{DC:HyperLogLogTheAnalysisOfANearoptimalCardinalityEstimationAlgorithm} and share with different hosts\cite{TON2017_CardinalityEstimationElephantFlowsACompactSolutionBasedVirtualRegisterSharing} to design a novel algorithm VBDR which can estimate several hosts' cardinalities under sliding time window. The main contributions of this paper is listed below: 
\begin{enumerate}
\item Propose a novel counter, Bit Distance Recorder(BDR), to record the state of the host under a sliding time window;
\item Propose a cardinalities estimating algorithm under a sliding time window, virtual Bit Distance Recorder VBDR. VBDR estimates different hosts' cardinalities with a fixed size of memory.   
\item Deploy VBDR on GPU to estimate cardinalities in real time; 
\end{enumerate}

This paper is arranged as below. In the next section, we will introduce the related works. In section 3, we will describe how BDR works and why it can maintain the most left ``1" bit position under sliding time window. Based on BDR, algorithm VBDR is proposed in section 4. Section 4 also introduces a method to deploy VBDR on GPU.  
\section{Background \& Related work}
\subsection{Problem definition}
Measuring the core network's properties, such as traffic size, packets number, host cardinality and so on, is the foundation of network management. This paper focuses on how to estimate different hosts' cardinalities over sliding time window. Suppose there is a core network, $ANet$, which is under the management of some organizations, institutes or ISP(internet service provider). $ANet$ communicates with other networks, denoted as $BNet$, through a set of edge routers ER. For a host $aip \in ANet$, its cardinality is the number of hosts in $BNet$ which communicate with $aip$ through $ER$ is a time window. The managers of $ANet$ have the authority to inspect every packet between $ANet$ and $BNet$ through ER.  So the task of cardinality estimation is to estimate hosts' cardinalities by scanning all packets passing through $ER$. 
\subsection{Cardinality estimation}
For a host $aip \in ANet$, let $Pkt(aip, t, k)$ represent the stream of packets passing through $ER$ in time window $W(t, k)$ whose source or destination IP address is $aip$. An IP pair which is similar to $<aip,bip>$ could be extracted from each packet in $Pkt(aip, t, k)$ where $bip$ is the other host in the packet. We also call $bip$ the opposite host of $aip$. Let $IPair(aip, t, k)$ represent to the stream of IP pairs corresponding to $Pkt(aip, t, k)$. Because a host $bip \in BNet$ could send several packets to or receive several packets from $aip$ in a time window, IP pair $<aip, bip>$ can appear many times in $IPair(aip, t, k)$. The number of distinct IP pairs in $IPair(aip, t, k)$ is the cardinality of $aip$. Let $OP(aip, t, k)$ represent the set of hosts in $BNet$ that communicate with $aip$ in W(t,k) and $|OP(aip)|$ represent the number of hosts in $OP(aip, t, k)$. Estimating the cardinality of $aip$ in $W(t, k)$ is to calculate $|OP(aip, t, k)|$ by scanning $IPair(aip, t, k)$. 

Many cardinality estimation algorithms have been proposed. Cardinality estimation algorithms use fix number of the counter to record and calculate the cardinality of a host. All these algorithms use a counter vector containing $g$ counters. What is preserved in a counter, how to update counters and how to estimate the cardinality from the counter vector are unique in different algorithms.

Flajolet et al. \cite{PCSA:ProbabilisticCountingAlgorithmsForDataBaseApplications}firstly proposed such an algorithm which is called Probabilistic Counting with Stochastic Averaging, PCSA. Each counter in PCSA is a bitmap containing 32 bits. For every opposite host of $aip$, a random selecting counter is used to record the least significant bit of this element. Least significant bit, LSB, is the first `1' bit starting from the right. After scanning all items in the stream, the value of each counter is its least zero position starting from the right. Cardinality could be acquired according to the sum of every counter. Scheuermann et al. proposed a more accuracy estimating equation when the load factor is smaller than 20. Load factor is the ratio of cardinality to $g$. 

The task of every counter in PCSA is to record the lowest zero position of every element. For an IPv4 address, the biggest value of least zero position is 32. But PCSA uses 32 bits to record the least zero position which leaves great improvement space. Because the biggest value of each counter is 32, 5 bits are big enough to represent it. Motivated by this idea, Philippe et al. proposed the LogLog counting algorithm\cite{DC:LoglogCountingOfLargeCardinalitiesDurand2003}. Unlike PCSA, each counter of LogLog records the leftmost `1' bit position of every element in the stream. Loglog estimates the cardinality according to the geometric mean value of all counters. 
Many algorithms are derived from LogLog. Flajolet et al. \cite{DC:HyperLogLogTheAnalysisOfANearoptimalCardinalityEstimationAlgorithm} found that when using the harmonic mean value of all the counters, the accuracy will be improved. And their proposed HyperLogLog algorithm based on this idea. HyperLogLog is the most memory efficient algorithms for large cardinality estimation. But the traditional HyperLogLog only works under discrete time window. Sliding time window has higher monitor resolution and many works have been done for cardinality estimation under sliding time window.

\subsection{Sliding time window vs. discrete time window}
 Discrete time window and sliding time window are two kinds of the period for cardinality estimating as shown in figure \ref{fig_sliding_and_discrete_time_window}. 

Traffic between network ANet and BNet could be divided into successive time slices which have the same duration. The length of a time slice could be 1 second, 1 minute or any period in different situations. A sliding time window, denoted as $W(t, k)$, contains k successive time slices starting from time point t as shown in the top part of figure \ref{fig_sliding_and_discrete_time_window}. Sliding time window will move forward one time slice a time. So two adjacent sliding time windows contain k-1 same slices. When k is set to 1, there is no duplicate period between two adjacent windows, which is the case of the discrete time window in the bottom part of figure \ref{fig_sliding_and_discrete_time_window}. In figure \ref{fig_sliding_and_discrete_time_window}, the size of the time slice is set to 1 second for sliding time window and 300 seconds for the discrete time window. A sliding window in figure \ref{fig_sliding_and_discrete_time_window} contains 300 time slices. In figure \ref{fig_sliding_and_discrete_time_window}, the size of a sliding time window is equal to that of a discrete time window. 

Cardinality estimation under discrete time window is straightforward because it doesn't need to maintain the appearance of opposite hosts in another time window. But the result is affected by the starting of the discrete time window. When a super point has different opposite hosts in two adjacent time windows, it may be neglected under discrete time window. 
 
Sliding time window has higher accuracy than discrete time window because it monitors traffic in a much more scalable way\cite{SDC2010SHLL:SlidingHyperLogLogEstimatingCardinalityDataStreamOverSlidingWindow}. 
 
Being required to preserve the state of opposite hosts in previous time slices, cardinality estimation under a sliding time window is more burdensome. But many works have been down trying to solve this problem. The main idea is to replace each counter used in a discrete time window with a more powerful structure which can tell if itself is active in the current time window. For a counter, if it is updated in $W(t,k)$, it is called active in this time window.

Fusy et al. \cite{SDC2007:EstimatingNumberActiveFlowsDataStreamOverSlidingWindow} extended MinCount to sliding window by maintaining a list of hosts that may become a minimum in a future window. The new algorithm is called Sliding MinCount. The minimum host is the latest arrived hosts among the set of hosts whose hashed value realizes the minimum in a sliding time window. When the time window sliding, Sliding MinCount updates every list and removes inactive hosts from these hosts list. But Sliding MinCount requires much space to store the minimum value of different time slices. In the worst case, each counter will maintain k minimum values in a sliding time window with k time slices. When using 32 bits to represent a minimum amount, each counter of Sliding MinCount requires 32*k bits.

Chabchoub et al. \cite{SDC2010SHLL:SlidingHyperLogLogEstimatingCardinalityDataStreamOverSlidingWindow} replaced each counter in HyperLogLog with a list of future possible maxima(LFPM). Each cell of LFPM uses 4 bytes to store timestamp and 1 byte to store the max leftmost 1-bit. In a sliding time window with k time slices, LFPM contains $ln(n/g)$ cells on average where n is the cardinality. So the size of a LFPM is $40*ln(n/g)$ bits. Denote this algorithm as LFPM-HLL. The size of LFPM-HLL is variable because every LFPM's size is not fixed. And LFPM-HLL is expected to consume g*40*log2(n/g) bits of memory on average. Dynamic memory is a burden for running time especially for parallel platform such as GPU. LFPM's purpose is to store the maxima left "1" bit position in the current time window. In this paper we propose a fixed size counter, bit distance recorder to solve this problem with smaller memory consumption.

\subsection{Multi hosts cardinality estimation}
\begin{figure}[!ht]
\centering
\includegraphics[width=0.47\textwidth]{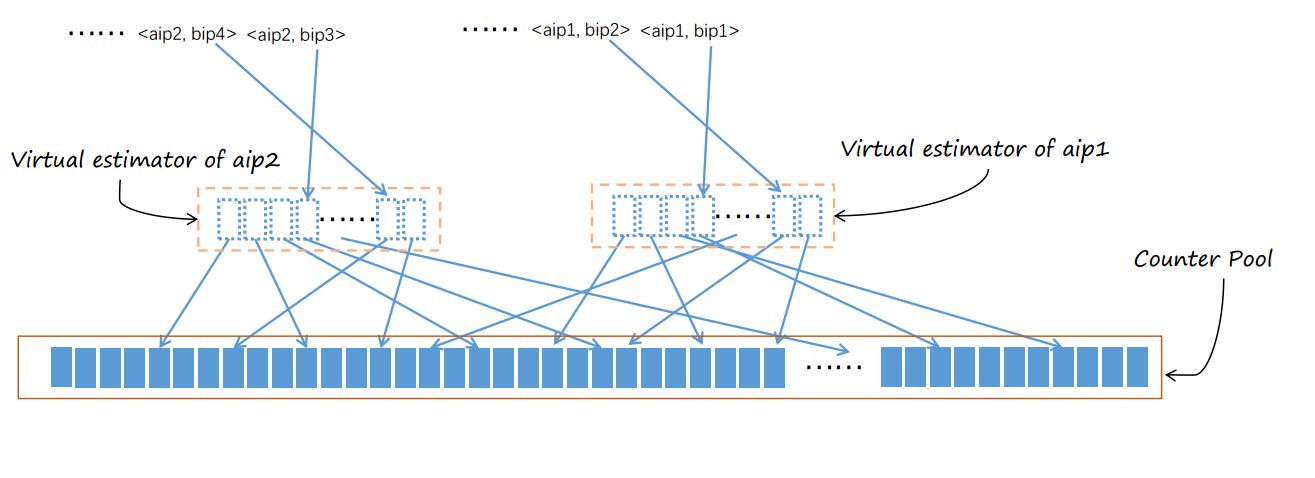}
\caption{Virtual counter vector}
\label{fig_virtural_counter_vector_illustrate}
\end{figure}

In the core network, there are huge hosts. A precisely way to acquire all of these hosts' cardinalities is to allocate an estimator for each of them. But this way is memory wasting and slow. Recent algorithms use a fixed number of estimators to maintain and calculate all hosts' cardinalities. Virtual estimator algorithm is one of the popular multi cardinalities estimation algorithms. Every host's logical counter vector shares counter with others in a counter pool as shown in fig.\ref{fig_virtural_counter_vector_illustrate}. The estimator of a virtual estimator could be LE \cite{HSD:SpreaderClassificationBasedOnOptimalDynamicBitSharing}\cite{HSD:GPU:2014:AGrandSpreadEstimatorUsingGPU}, HyperLogLog\cite{TON2017_CardinalityEstimationElephantFlowsACompactSolutionBasedVirtualRegisterSharing}. Virtual counter based algorithms only need to update a counter for a packet and have faster packets scanning speed.
Every host has a virtual counter vector to estimate its cardinality. Every counter in the virtual counter vector is relative to a physical counter in the pool. A physical counter is shared by several virtual counter vector. Counters in \cite{TON2017_CardinalityEstimationElephantFlowsACompactSolutionBasedVirtualRegisterSharing} records the biggest left 1 bits position under discrete time window. In this paper we change every counter in it with a powerful counter to let it work under sliding time window. 

\section{Bit distance recorder}
Under a discrete time window, a counter with $log_2(log_2(n/g))$ bits is big enough to store the most left ``1" bit position where n is the number of distinct opposite hosts. Let LBP1 represent the most left ``1" bit position as defined in HyperLogLog. But under sliding time window, the value of this counter may become inactive when the window slides forward. The critical step to maintain the LBP1 under sliding time window is to determine it is active in the current time window. In this section, we will introduce a novel counter, Bit distance recorder($BDR$), to solve this problem.

Suppose a sliding time window contains k time slices at most. Let BDR contains two part: current time slice LBP1(denoted as nowLBP1) and a vector consisting of $ceil(log_2(n/g))$ distance recorders(denoted as DRV). A distance recorder, written as DR, is z bits where z is an integer at least $ceil(log_2(k+1))$ bits. LBP1 has $ceil(log_2(n/g))$ different values, and one value is corresponding to a DR. Let $DRV[i]$ represent the DR of LBP with value $i$. DR is designed to indicate if a LBP is active in the current time window. A DR has the following operations:
\begin{enumerate}
\item $InitDR(dr)$: set the value of "$dr$" to $2^z -1 $. This operation initializes an $dr$ at the beginning. 
\item $SetDR(dr)$: set the value of "$dr$" to 0.
\item $SlideDR(dr)$: if $dr \leq 2^k-1$, $dr++$.
\item $IsActiveDR(dr)$: return if $dr$ is active. if $dr < k$, $dr$ is active and return true, else $dr$ is inactive and return false.
\end{enumerate}

For a BDR, let $ST'(t)$ represent the stream of hashed opposite hosts that mapped to it in the time slice $t$. The hashed opposite host is the result of a hash function as defined in HyperLogLog to make sure that the opposite hosts are distributed uniformly. While scanning $ST'(t)$, nowLBP1 stores the LBP1 of $ST'(t)$. After scanning all elements in $ST'(t)$, BDR is updated by EndSliceUpdateBDR() as show in algorithm \ref{alg-updateBDR_after_scanningHosts}.
\begin{algorithm}                       
\caption{EndSliceUpdateBDR}    
\label{alg-updateBDR_after_scanningHosts}                     
\begin{algorithmic}
\Require{ BDR }
\Comment{\\ //BDR.DRV represent the DRV of BDR }
\For dr in BDR.DRV 
\State SlideDR(dr)
\EndFor
\Comment{\\ //BDR.nowLBP1 represent the nowLBP1 of BDR }
\State SetDR(BDR.DRV[BDR.nowLBP1])
\State Return
\end{algorithmic}
\end{algorithm}

After being updated, the LBP1 used for cardinality estimation could be acquired by algorithm GetLBP1BDR() as show below.
\begin{algorithm}                       
\caption{GetLBP1BDR}    
\label{alg-GetLBP1BDR}                     
\begin{algorithmic}
\Require{ BDR }
\Ensure{LBP1}
\Comment{\\ //BDR.DRV represent the DRV of BDR }
\State $lbp1 \Leftarrow log_2(n/g)$
\While {$lbp1 \geq 0$}
\If { IsActiveDR(BDR.DRV[lbp1])}
  \If{ $ BDR.DRV[lbp1] < k$} \\
     \Return lbp1
   \EndIf
\EndIf
\State $lbp1--$
\EndWhile\\
\Return 0
\end{algorithmic}
\end{algorithm}

The ``nowLBP1"  member of BDR helps to reduce the memory writing because when the new LBP1 is smaller than $nowLBP1$, it won't be recorded. The ``DRV" member of BDR could also be updated directly while scanning IP pair without using $nowLBP1$, and this method will be discussed in the following section to handle IP pair parallel in GPU.

GetLBP1BDR(BDR) acquires the value of BDR by scanning the most active LBP1 in DRV. After acquire LBP1 in the current time window, cardinality could be estimated like the HyperLogLog. The next section will introduce the VBDR which estimates several hosts' cardinalities at the same time.

\section{Sliding cardinalities estimation}
For the sake of brevity, we suppose there are two networks, $ANet$ and $BNet$. All traffic between $ANet$ and $BNet$ are transmitted through an edge router $ER$. And we want to calculate the cardinalities of hosts in $ANet$. From the perspective of $ER$, the cardinality of a host is defined as below. 

\begin{definition}{Host cardinality}
\label{def_hostCardinality}
For a host $aip \in ANet$, its cardinality in time window $W(t, k)$ is the number of hosts in $BN$ that send packets to or receive packets from it through $ER$ in $W(t, k)$, written as $|OP(aip, t, k)|$.
\end{definition}
This paper uses a novel algorithm, Virtual Bit Distance Recorder(VBDR), to estimate different hosts' cardinalities under a sliding time window. VBDR is derived from HyperLogLog and Virtual Estimator. 
\subsection{Virtual Bit Distance Recorder}
$VBDR$ contains a pool of $BDR$ denoted as $BDRP$. There are total $z$ $BDR$ in $BDRP$ and let $BDRP[i]$ represent the $i$-th $BDR$ in it. For every host ``aip" in $ANet$, there are $g$ different $BDR$ selected from $BDRP$ corresponding with it, and $g=2^b$ where $b$ is a positive integer smaller than 32. Let $VBV(aip)$ represent these $g$ $BDR$ in $BDRP$ and $VBV(aip)[i]$ represent the $i$-th $BDR$ in $VBV(aip)$. $VBV(aip)[i]$ is related with a physical $BDR$ in $BDRP$. Let $H(x,N,A)$ represent a random hash function with seed $A$ that maps an integer x to an integer smaller than $N$. The index of physical $BDR$ of $VBV(aip)[i]$ could be acquired by algorithm \ref{alg-getPhysicalBDR}.

\begin{algorithm}                       
\caption{getPhyIdx}    
\label{alg-getPhysicalBDR}                     
\begin{algorithmic}     

\Require{ Host IP: aip\\
          Virtual index: i \\
          Hash function seed: $A_0$}
\Ensure{Physical index: j}
\State $s_1 \Leftarrow H(i,2^{32},A_0)$
\State $j \Leftarrow H(aip,z,s_1)$\\
\Return $j$
 
\end{algorithmic}
\end{algorithm}

Every $BDR$ in $BDRP$ is shared with different hosts in $ANet$. An IP pair, like $<aip,bip>$ where $aip \in ANet$ and $bip \in BNet$, could be extracted from every packet between $ANet$ and $BNet$. Let $IPpiar(t)$ represent the stream of IP pair in time slice t. $VBDR$ scans every IP pair in $IPpair(t)$ as shown in algorithm \ref{alg-scanIPpair}. Let $LB(x,i)$ return the left $i$ bits of the binary form of integer $x$.

\begin{algorithm}                       
\caption{scanIPpair}    
\label{alg-scanIPpair}                    
\begin{algorithmic}     

\Require{ IP pair stream in time slice t: $IPpair(t)$\\
          Hash function seed: $A_0$, $A_1$ }
\For {$<aip,bip> \in IPpair(t)$}
\State $bip' \Leftarrow H(bip,2^{32},A_1)$
\State $vidx \Leftarrow LB(bip',b)$
\State $bip' \Leftarrow bip' << b$
\State $pidx \Leftarrow getPhyIdx(aip,vidx,A_0)$
\State $BDRP[pidx].nowLBP1 \Leftarrow max(BDRP[pidx].nowLBP1, LBP1(bip'))$
\EndFor

\For {$bdr \in BDRP$}
\State EndSliceUpdateBDR(bdr)
\EndFor \\
\Return
 
\end{algorithmic}
\end{algorithm}

After scanning all IP pairs in $IPpair(t)$, $BDRP$ stores the latest hosts' LBP1. Cardinalities could be estimated from $BDRP$. According to paper \cite{TON2017_CardinalityEstimationElephantFlowsACompactSolutionBasedVirtualRegisterSharing}, to estimate a host's cardinality we should get the sum of LBP1. For a host ``aip" in $ANet$, its sum of LBP1 in $W(t,k)$ could be acquired by algorithm \ref{alg_estimateCHLL} after processing $IPpair(t+k-1)$ by algorithm \ref{alg-scanIPpair}.

\begin{algorithm}                       
\caption{getSumLBP1}    
\label{alg_estimateCHLL}                    
\begin{algorithmic}     

\Require{ Host IP: $aip$\\
          Hash function seed: $A_0$}
\Ensure {The sum of LBP1 of $aip$}
\State $sLBP1 \leftarrow 0$
\For {$ i \in [0,g-1]$}
\State $pidx \Leftarrow getPhyIdx(aip,i,A_0)$
\State $sLBP1 \Leftarrow sLBP1 + GetLBP1BDR(BDRP[pidx])$
\EndFor \\
 \Return $sLBP1$
 
\end{algorithmic}
\end{algorithm}
After getting the sum LBP1 of ``aip" we could get its cardinality by equation (5) in paper \cite{TON2017_CardinalityEstimationElephantFlowsACompactSolutionBasedVirtualRegisterSharing}. Noted that, $BDRP$ could also be used in PCSA\cite{PCSA:ProbabilisticCountingAlgorithmsForDataBaseApplications}, LogLogcount\cite{RSL2012:TheLogLogCountingReversibleSketchADistributedArchitectureForDetectingAnomaliesBackboneNetworks} by replacing its nowLBP1 with the value recorded in other algorithms. VBDR is also suitable for parallel running after a few modification. In the next section we will introduce how to run it on a famous parallel platform GPU.

\subsection{Running on GPU}
VBDR can also run on GPU. Only IP pair of a packet is necessary and transmit only IP pairs to GPU is a more efficient way. 
In a high-speed network, such as 40 Gb/s, there are millions of packets passing through the edge of the network. To scan so many packets in real time requires plenty of computing resource. CPU is one of the most general computing parts, and each core of it is potent to deal with complex tasks running different instructions. Though a core in the CPU is powerful, its price is very high. If we want to use hundreds of CPU cores to deal with high-speed traffic parallel, we have to generate a cluster with several CPUs. The cost of the cluster will be increased with its scale. Graphics processing unit (GPU) is one of the most popular parallel computing platforms in recent years. GPU contains hundreds of processing unit in a chip, much more than that CPU has. For these tasks that have no data accessing conflict and processing different data with the same instructions (SIMD), GPU can acquire a high-speed up\cite{PD2013:BenchmarkingOfCommunicationTechniquesForGPUs}\cite{PD2013:GeneratingDataTransfersForDistributedGPUParallelPrograms}. Every packet is processed by algorithm \ref{alg-scanIPpair}. 

Algorithm \ref{alg-scanIPpair} scans every IP pair and sets the nowLBP1 of BDR to the biggest value. Under parallel environment, reading-writing conflict may arise because of parallel updating. For example, two threads, $thd_1$ and $thd_2$, are scanning two different opposite hosts, $bip_1$ and $bip_2$, at the same time. Because a counter is sharing by several host, these two threads may updating the same physical BDR. Suppose these two threads are updating the $i$-th $BDR$ in $BDRP$ and $r_0=BDRP[i].nowLBP1$. Let $r_1=LBP1((H(bip_1,2^{32},A_0)<< b))$ and $r_2=LBP1((H(bip_2,2^{32},A_0)<< b))$. If $r_0 < r_1 < r_2$, $BDRP[i].nowLBP1$ maybe set to $r_1$ by mistake. Because when $thd_2$ reads $r_0$ it will rewrite $BDRP[i].nowLBP1$ with $r_2$. If $thd_1$ read $r_0$ before $thd_2$ rewrite it, $thd_2$ will rewrite $BDRP[i].nowLBP1$ with $r_1$. If the the rewriting task of $thd_1$ is submitted to the memory after that of $thd_2$, $r_1$ will cover $r_2$. But according to algorithm \ref{alg-scanIPpair} under serial process, $BDRP[i].nowLBP1$ should be $r_2$ after $thd_1$ and $thd_2$ finishing. To over come this problem, we replace nowLBP1 of every BDR with a bit string with size $32-b$, denoted as bsLBP1. Every bit of bsLBP1 corresponds with a LBP1 and let bsLBP1[i] represent the $i$-th bit in bsLBP1. Every bit of bsLBP1 is reset to 0 at the beginning of every time slice. When scanning IP pair, if $LBP1(bip')=r$, then set $bsLBP1[r]=1$. The LBP1 of a DR at a time slice is acquired by finding the biggest ``1" bit position.
The BDR updating algorithm at the end of a time slice on GPU, EndSliceUpdateBDRGpu(), is show in algorithm \ref{alg-updateBDR_after_scanningHosts_GPU}.
\begin{algorithm}                       
\caption{EndSliceUpdateBDRGpu}    
\label{alg-updateBDR_after_scanningHosts_GPU}                     
\begin{algorithmic}
\Require{ BDR }
\Comment{\\ //BDR.DRV represent the DRV of BDR }
\For {dr in BDR.DRV }
\State SlideDR(dr)
\EndFor
\Comment{\\ //BDR.bsLBP1 represent the bsLBP1 of BDR }
\State $lbp1 \leftarrow 0$
\For {$i \in [0,32-g-1]$}
  \If {$BDR.bsLBP1[i] == 0$}
    \State $lbp1 \leftarrow i$
   \EndIf
 \EndFor
\State SetDR(BDR.DRV[lbp1])
\State Return
\end{algorithmic}
\end{algorithm}

Because algorithm \ref{alg_estimateCHLL} only read BDV, so it is not affected by the parallel running. A bit could be set by several threads at the same time, so BDR with bsLBP1 could be running on GPU to scan several IP pair parallel. Algorithm \ref{alg-scanIPpairGpu_part1} shows how to scan IP pair on GPU. Every thread of GPU reads a IP pair from $IPpair(t)$ and running algorithm \ref{alg-scanIPpairGpu_part1} to handle it. Thousands of threads on GPU handle the same number of IP pairs at the same time. After scanning all IP pairs in $IPpair(t)$, every BDR will be updated by a GPU thread with algorithm \ref{alg-updateBDR_after_scanningHosts_GPU}. Thousands of threads on GPU handle the same number of BDR at the same time. So no matter IP pair scanning or BDR updating, GPU can finished more quickly than that under CPU.

\begin{algorithm}                       
\caption{scanIPpairOnGpu}    
\label{alg-scanIPpairGpu_part1}                    
\begin{algorithmic}     

\Require{ An IP pair: $<aip, bip>$\\
          Hash function seed: $A_0$, $A_1$ }

\State $bip' \Leftarrow H(bip,2^{32},A_1)$
\State $vidx \Leftarrow LB(bip',b)$
\State $bip' \Leftarrow bip' << b$
\State $pidx \Leftarrow getPhyIdx(aip,vidx,A_0)$
\State $BDRP[pidx].bsLBP1[LBP1(bip')] = 1$

\end{algorithmic}
\end{algorithm}

Algorithm \ref{alg-scanIPpairGpu_part1} uses a $bsLBP1$ to recorder the presentation a LBP1 in the current time slice. Because the $bsLBP1$ is only updated by bit-wise operations, exactly to speak is the ``bit-wise OR" operation, so it has a fast speed. But $bsLBP1$ requires at least $log_2(n/g)$ bits which increases the memory requirement. As mentioned before, $DRV$ could be used to record the LBP1 directly, and $nowLBP1$ could be removed. Algorithm \ref{alg-scanIPpairGpu_part2} shows how to merely use DRV while scanning IP pair. When we use $DRV$ to record the LBP1 of every IP pair's LBP1, $DR$ preserves the latest $LBP1$. But the max $LBP1$ could be acquired from $DRV$ by algorithm \ref{alg-GetLBP1BDR} too. When $BDR$ only contains the DRV, preserving the state of BDR after scanning all IP pairs in $IPpair(t)$ is simple as shown in algorithm \ref{alg-updateBDROnlyDRV_before_scanningHosts_GPU}.But algorithm \ref{alg-updateBDROnlyDRV_before_scanningHosts_GPU} should be done at the beginning of a time slice before scanning the IP pairs. For a time slice, before scanning the IP pairs in it, every BDR will be updated by a GPU thread with algorithm \ref{alg-updateBDROnlyDRV_before_scanningHosts_GPU}. Thousands of threads on GPU handle the same number of BDR at the same time. 

\begin{algorithm}                       
\caption{BeginSliceUpdateDRV}    
\label{alg-updateBDROnlyDRV_before_scanningHosts_GPU}                     
\begin{algorithmic}
\Require{ BDR }
\Comment{\\ //BDR.DRV represent the DRV of BDR }
\For { $dr \in BDR.DRV$ }
\State SlideDR(dr)
\EndFor
\State Return
\end{algorithmic}
\end{algorithm}

\begin{algorithm}                       
\caption{scanIPpairUpdatDRV}    
\label{alg-scanIPpairGpu_part2}                    
\begin{algorithmic}     

\Require{ An IP pair: $<aip, bip>$\\
          Hash function seed: $A_0$, $A_1$ }

\State $bip' \Leftarrow H(bip,2^{32},A_1)$
\State $vidx \Leftarrow LB(bip',b)$
\State $bip' \Leftarrow bip' << b$
\State $pidx \Leftarrow getPhyIdx(aip,vidx,A_0)$
\State $SetDR(BDRP[pidx].DRV[LBP1(bip')])$

\end{algorithmic}
\end{algorithm}

Algorithm \ref{alg-scanIPpairGpu_part2} only sets $DR$ while scanning IP pairs. The $SetDR()$ operation just set a $DR$ to zero and this operation could be launched by different threads parallel. So algorithm \ref{alg-scanIPpairGpu_part2} could be launched by different threads of GPU parallel. No matter algorithm \ref{alg-scanIPpairGpu_part1} or algorithm \ref{alg-scanIPpairGpu_part2}, their just effect the IP pair scanning process. After scanning all IP pairs in $IPpair(t)$ by one of these algorithms, the LBP1 of BDR could be acquired by algorithm \ref{alg-GetLBP1BDR}. So these algorithms can use the same cardinality estimation method as mentioned before.

Not only IP pairs scanning but also hosts' cardinalities estimation can run on GPU. A BDR could be read by several threads without conflict. So several hosts' cardinalities could be estimated at the same time.

From the discussion before, we see that VBDR has three versions: serial running VBDR (VBDR-serial), parallel running VBDR with high speed on GPU($VBDR-gfast$) and parallel running VBDR with small memory requirement on GPU(VBDR-gsmall). The memory consumption of BDR of these versions are listed in table \ref{tbl-vbdr-memory}.
\begin{table*}
\caption{Memory consumption of different kinds of VBDR}
\label{tbl-vbdr-memory}
\begin{tabular}{c|c}
\hline 
Algorithm & Size of a BDR (bits) \\ 
\hline 
VBDR-serial & $log_2(log_2(n/g))+log_2(n/g)*log_2(k+1)$ \\ 
VBDR-gfast & $log_2(n/g)+log_2(n/g)*log_2(k+1)$ \\ 
VBDR-gsmall & $log_2(n/g)*log_2(k+1)$ \\ 
\hline 
\end{tabular} 
\end{table*}
Although VBDR-gfast consumes the most memory among these VBDR algorithms, but it has the fastest IP pair scanning speed. VBDR-serial only works under serial environment such as on a single CPU thread. VBDR-gfast and VBDR-gsmall can works under both serial and parallel environment, but their IP pair scanning speed or memory requirement is not as good as that of VBDR-serial. All of these algorithms have their merits and weaknesses. Users can choose the suitable one for different situations. For example, if the platform has no GPU card, VBDR-serial is the only choice. If the server has a low level GPU card, such as GTX 650 with only 1 GB graphical memory, VBDR-gsmall is a better choice because it can save memory. If the server has an advanced GPU card, such as Nvidia Titan XP with 12 GB memory, users can choose VBDR-gfast to acquire the fast speed.

\section{Conclusion}
VBDR is a fast and memory efficient cardinalities estimating algorithm under a sliding time window. It uses BDR to preserve the LBP1 of every time slice. With BDR, memory efficient cardinality estimating algorithms, such as PCSA, LogLogCount, HyperLogLog and so on, could be extended to sliding window environment. VBDR is such one that uses BDR as counter of virtual estimator to estimate several hosts' cardinalities at the same time. VBDR estimates every host's cardinality with a virtual BDR vector from a BDR pool. The more BDR in the pool, the higher accuracy of VBDR will be. In this paper we also improve the performance of VBDR, and propose three versions of VBDR to estimate cardinalities under serial environment and parallel environment. Users can choose one from these three versions for different environments.

\iftoggle{ACM}{
\bibliographystyle{ACM-Reference-Format}
}
\iftoggle{IEEEcls}{
\bibliographystyle{IEEEtran}
}
\iftoggle{ElsJ}{
\bibliographystyle{elsarticle-num}
}

\bibliography{..//ref} 

\end{document}